# Revisiting the Sample Adaptive Offset post-filter of VVC with Neural-Networks

Philippe Bordes
*InterDigital*
Cesson-Sevigné, France
philippe.bordes@interdigital.com

Franck Galpin
*InterDigital*
Cesson-Sevigné, France
franck.galpin@interdigital.com

Thierry Dumas
*InterDigital*
Cesson-Sevigné, France
thierry.dumas@interdigital.com

Pavel Nikitin
*InterDigital*
Cesson-Sevigné, France
pavel.nikitin@interdigital.com

*Abstract*—The Sample Adaptive Offset (SAO) filter has been introduced in HEVC to reduce general coding and banding artefacts in the reconstructed pictures, in complement to the De-Blocking Filter (DBF) which reduces artifacts at block boundaries specifically. The new video compression standard Versatile Video Coding (VVC) reduces the BD-rate by about 36% at the same reconstruction quality compared to HEVC. It implements an additional new in-loop Adaptive Loop Filter (ALF) on top of the DBF and the SAO filter, the latter remaining unchanged compared to HEVC. However, the relative performance of SAO in VVC has been lowered significantly. In this paper, it is proposed to revisit the SAO filter using Neural Networks (NN). The general principles of the SAO are kept, but the a-priori classification of SAO is replaced with a set of neural networks that determine which reconstructed samples should be corrected and in which proportion. Similarly to the original SAO, some parameters are determined at the encoder side and encoded per CTU. The average BD-rate gain of the proposed SAO improves VVC by at least 2.3% in Random Access while the overall complexity is kept relatively small compared to other NN-based methods.

## I. Introduction

The state-of-the-art block-based hybrid video coding frameworks introduce some blocking and ringing artifacts. To attenuate these artifacts, the last video coding standards have designed various in-loop restoration tools. For example, H.264/AVC defined a deblocking filter and H.265/HEVC developed an additional Sample Adaptive Offset (SAO) filter. In H.266/VVC, three in-loop restoration filters are cascaded: a revisited version of the deblocking filter [3], the SAO which is identical to HEVC [4] and a Wiener-based Adaptive Loop Filter (ALF) [5]. Deblocking filter is mainly used to attenuate blocking artifacts. SAO filter mainly filters ringing artifacts and color biases. ALF tries to minimize the mean square error between original samples and reconstructed samples using Wiener-based adaptive filter coefficients.

VVC improves the compression rate by about 36% [7] in Random Access (RA) Common Test Conditions (CTC) [8] compared to HEVC using the objective BD-rate PSNR metric [9]. This level of performance is achieved with several new coding tools, but their relative performance may be quite different. For example, the BD-rate gain provided by SAO and ALF is about 0.1% and 4.3% respectively in RA for the luma component with VVC [10], [11], whereas it was about 2% for SAO with HEVC [12].

At the same time, recent works with Convolutional Neural Networks (CNNs) for image restoration have shown that one can significantly improve the codec performance with NN-based post-filtering. For example, the JVET group recently created a group for studying how the NN technology may further improve VVC [13]. However, most of the previous works leverage on powerful GPUs platforms and the amount of calculation per pixel is far beyond the capacity and power of consumer or mobile devices for the decade, even considering the natural evolution of the processors technology [14].

Both the relative low performance of SAO in VVC and the recent developments in CNNs for image restoration have motivated the study of this NN-based post-filter targeting relatively small complexity.

The paper is organized as follows: in Section II, the SAO in-loop post-filter is presented. The proposed framework and the NN architecture are described in Section III. Next, the implementation with VVC and the experimental results are depicted and discussed. Finally, we conclude and present future perspectives.

## II. Sample adaptive offset

The concept of SAO is to reduce mean sample distortion of a region (a.k.a Coding Tree Unit (CTU)) by first classifying the samples into multiple categories with a selected classifier, obtaining an offset for each category, and then adding the offset to each sample of the category, where the selected classifier index and the offsets of the region are coded in the bitstream. There are six SAO types and classes which are selectable per CTU and per component type (Luma, Chroma): off (OFF), edge offset (EO_0, EO_90, EO_135, EO_45) and band offset (BO).

In the case of an EO classifier, each reconstructed sample is classified into NC = 5 categories, depending on the local gradients along the direction given by this EO classifier, as depicted in Fig.1. (NC - 1) offset values are coded, one for each category (one category has offset equal to zero).

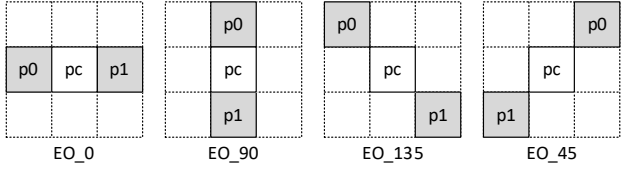

| Category | Condition | Meaning |
|---|---|---|
| 1 | $p_c < p_0$ and $p_c < p_1$ | local mean |
| 2 | $p_c < p_0$ and $p_c == p_1$ or $p_c == p_0$ and $p_c < p_1$ | edge |
| 3 | $p_c > p_0$ and $p_c == p_1$ or $p_c == p_0$ and $p_c > p_1$ | edge |
| 4 | $p_c > p_0$ and $p_c > p_1$ | local max |
| 0 | None of the above | plain |

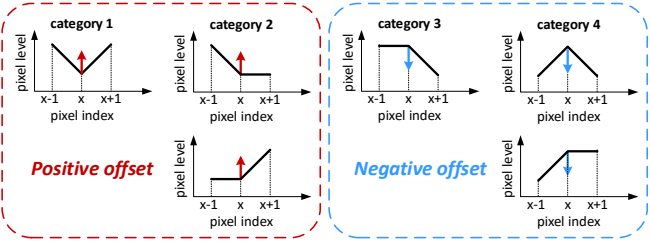

Fig. 1. EO classifiers categories. In the above table, $p_c$ is the current sample value. $p_0$ and $p_1$ are the values of the previous sample and the next sample respectively along the direction given by the selected EO classifier.

In case of BO classifier, the pixel range of values (ex: 0..255, in 8-bit) is uniformly divided into 32 bands. The sample values belonging to (NC - 1) = 4 consecutive bands are modified by adding an offset *off(n)* as depicted in Fig.2. *nbOffsets* = (NC - 1) offset values are coded, one for each of the (NC - 1) consecutive bands (the remaining bands has offset equal to zero), and the position of the first band of the (NC - 1) consecutive bands within the 32 bands is also coded in the bitstream.

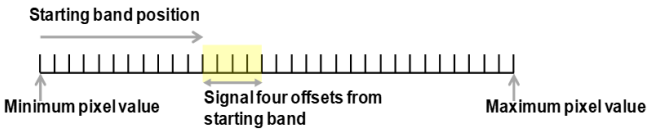

Fig. 2. BO classifier sends the offsets of four consecutive bands and the position of the first band of the (NC – 1) consecutive bands.

In case of OFF type, the reconstructed samples are not modified. The types and offsets for the CTU are possibly not coded but copied from the neighboring above or left CTU in Merge mode.

## III. PROPOSED APPROACH

To some extent, the sample classification operated by EO of SAO, which is based on 3x3 regions around the reconstructed pixels, shares some analogy with convolutional neural networks. Then it may be fairly natural considering improving the regular SAO using CNNs, targeting relatively small complexity to reduce the gap with possible future deployment. Another motivation is to investigate a single framework that could address a wide range of video coding configurations and bit rates.

### A. Proposed framework

The overall proposed framework (SAO-CNN) is depicted in Fig.3. It keeps major features of the SAO. But, the selection of which classifier to employ is replaced with a selection among K CNNs. Whereas SAO classifies the reconstructed samples into NC classes, each CNNs outputs a weight *w(s)* for each sample *s* which will modulate the correction for the sample *s*. This empowers much more classes than for the NC classes of SAO since all different output values of the CNN may be considered as a different class potentially.

Also, similarly to SAO, the additive correction is controlled with some transmitted offset parameters $off_i$ associated with any CNN (i) used in the current CTU:

$$corr(s) = w(s).off_i \qquad (1)$$

However, since the CNN output provides implicit classification, the number of different corrections values *corr* is substantially greater than the 4 offsets values of SAO.

To add even more flexibility to the proposed SAO-CNN, we provided the framework with the capability to combine the output of several (*M*) CNNs as proposed in [15]:

$$corr(s) = \sum_{i=0}^{M-1} w(s)_i . off_i \qquad (2)$$

The combination of several CNN outputs leverages the number of correction patterns while can reduce the number of CNNs.

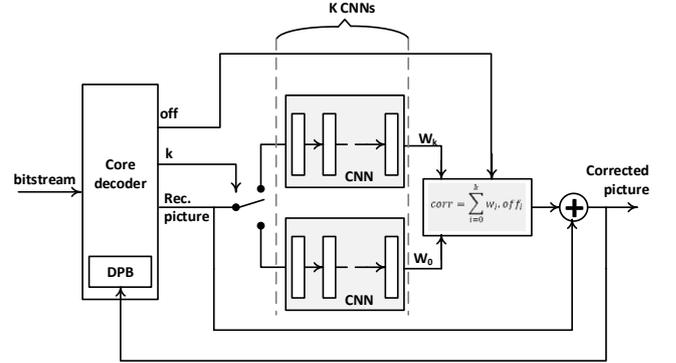

Fig. 3. The proposed SAO-CNN framework.

### B. Coding syntax

Given the similarity of the proposed filter with SAO, the coding of the CTU filter parameters re-uses most of the regular SAO syntax elements as shown in Table I. The *typeIdx* is used to code the index of the CNNs and the $off_i$ control parameters are coded with *absOff*. However, since the amount of correction tends to increase at low bit-rate and is correlated with the level of correction in the region, the values of $off_i$ are predicted with the neighboring CTUs.

TABLE I.   SYNTAX OF SAO IN VVC (LEFT) AND PROPOSED (RIGHT). C DENOTES THE INDEX OF THE COLOR COMPONENT. IN THE RIGHT COLUMNS, A BLACK CELL INDICATES THAT THE SYNTAX IN THE CELL ROW DOES NOT APPLY TO SAO IN VVC OR TO THE PROPOSED METHOD RESP.

|  | SAO | SAO-CNN |
|---|---|---|
| **merge_left** |  |  |
| if ( !merge_left ) **merge_up** |  |  |
| if ( !merge_left && !merge_up ) { |  |  |
| for c in {0,1,2} { |  |  |
| if (c < 2) **typeIdx0**[c] | OFF, EO, BO | OFF |
| if (typeIdx0[c] != OFF) { |  |  |
| for ( m=0..M ) |  | M=nbOffset |
| if (c < 2) **typeIdx1**[c][m] |  |  |
| for ( i=0..nbOffset ) **absOff**[i] | nbOffset=4 |  |
| if ( typeIdx0[c][0] == BO ) { |  |  |
| for ( i=0..nbOffset ) **signOff**[i] |  |  |

```
        bandOff
    }
    else {
        if (c < 2) eo_class
    }
  }
 }
}
```

## C. Networks architecture

The network architecture was chosen to reduce the overall complexity by minimizing the number of operations per pixels (MACs) and the number of parameters. Then, the CNNs are composed of a relatively small number $L$ of layers (6) and reduced number $n(l)$ of latents/channels per layer $l$ (16 to 32 or 64). Two versions v1 and v2 are provided, v2 being slightly less complex than v1 (Table II).

The proposed neural network architecture is depicted in Fig.4.

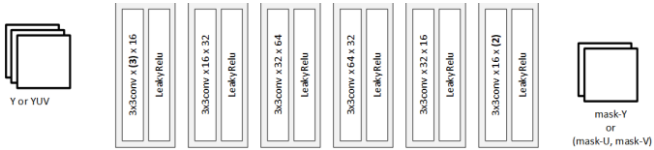

Fig. 4. Proposed NN architecture (v1).

Considering the CTU horizontal and vertical size is $S$ and for a convolution kernel of size $c \times c$, the MAC per pixel can be derived as follows:

$$MAC = \left(\frac{c.c}{S.S}\right)^2 . \sum_{l=0}^{L-1} n(l).n(l+1).(S + 2.(L-1))^2 \quad (3)$$

TABLE II.   NN PARAMETERS SUMMARY

| parameter | V1 | V2 |
|---|---|---|
| Number of layers L | 6 | |
| Num. input(s) (ni) | luma: 1 (Y), chroma 3 (YUV) | |
| Num. output(s) (no) | luma: 1 (Y), chroma 2 (UV) | |
| n(l) channels/latents | {ni,16,32,64,32,16,no} | {ni,16,16,32,16,16,no} |
| Conv kernel c x c | 3 x 3 | |
| Block size S x S | 128 x 128 | |
| MAC / pel (luma) | M x 50K | M x 15K |
| MAC / pel (chroma) | 50K | |

## IV. EXPERIMENTS AND RESULTS

### A. Training

In many previous works [15][16], the frameworks are de-facto designed for addressing a subset of possible video coding configurations (ex: intra or random-access) or bit-rate, since several separate models are trained and employed for different bit-rates.

The inner capability of SAO-CNN to combine multiple CNNs can soften the model transition and adapt to the temporal and spatial video sequence characteristics variability. Whereas some other previous works [15][17] were using single CNN with multiple outputs.

In our experiments, we employ 800 video sequences from the BVI-DVC dataset [18] to build our training database. Each sequence contains 64 frames, with 10 bit and YCbCr 4:2:0 format at four different resolutions from 270p to 2160p. These sequences were compressed by VVC VTM reference software codec using the JVET-CTC Random Access (RA) and All Intra (AI) configurations, with four QP values: 22, 27, 32 and 37. During encoding and decoding, the regular in-loop filters were kept active. For each QP, compressed video frames before SAO and their corresponding original counterparts were randomly selected, segmented into 128x128 image blocks, and converted to YCbCr 4:4:4 format by up-scaling the chroma.

The proposed CNN was implemented and trained using TensorFlow framework with the following training parameters: *l2* as loss function, Adam optimization, batch size of 20 and 130 training epochs. Sixteen models corresponding to intra(I)/inter(B) picture types, picture size CD/AB and QP 22-37 were trained (Table III):

TABLE III.   NN MODELS TRAINING FEATURE

| model | mode | Pic. size | QP | model | mode | Pic. size | QP |
|---|---|---|---|---|---|---|---|
| 0 | intra | AB | 22 | 8 | inter | AB | 22 |
| 1 | intra | AB | 27 | 9 | inter | AB | 27 |
| 2 | intra | AB | 32 | 10 | inter | AB | 32 |
| 3 | intra | AB | 37 | 11 | inter | AB | 37 |
| 4 | intra | CD | 22 | 12 | inter | CD | 22 |
| 5 | intra | CD | 27 | 13 | inter | CD | 27 |
| 6 | intra | CD | 32 | 14 | inter | CD | 32 |
| 7 | intra | CD | 37 | 15 | inter | CD | 37 |

### B. Implementation in VTM

The SAO-CNN framework has been implemented in place of the regular SAO in VTM-10. A module for inferring the CNN-models has been developed in C++, using 32-float or 16-bit quantized weights. In the latter case, the 32-float NN weights obtained from the training stage have been quantized into 16-bits so that all the calculations can be carried out in integers (16-bit latents and 32-bit accumulation), with saturation to avoid overflow. This allows benefiting of SIMD acceleration, speeding-up encoder and decoder by 1.4 and 2.1 respectively compared to 32-float model, while the BD-rate performance remains almost unchanged, as depicted in Table IV. Besides speeding-up inference, the integer calculation ensures strict bit-accuracy between the decoder and encoder reference pictures, so that no drift error occur in the reconstructed frames.

At the encoder side, the RDO algorithm which chooses the best parameters (selection of the NN model indexes and offset values) of the regular SAO, is maximally re-used (ex: re-use of same Lagrange parameters).

### C. Results

Tables V and VI summarize the compression performance of the SAO-CNN in-loop post-filtering module on the main common test sequences used in JVET and compared to the regular VTM-10 anchors. The BD-rates are computed with four QPs {22,27,32,37}. It can be observed that our proposed approach achieves significant and consistent coding gains on all test sequences and all configurations when integrated into VTM, with average BD-rate gains of -2.86% and -2.42% in Luma for Random Access and All Intra respectively, and about -7% and -5% in chroma, for v1 architecture.

The encoding and decoding times ratio have been obtained on a single core processor without GPU. Compared to VTM-10, the encoder complexity remains below 1.3x and the decoding time is between 6x and 34x for v2 architecture.

## D. Complexity analysis

In the Fig.5, the BD-rate gain versus number of MAC per pels (left) and the BD-rate gain versus the decoding time ratio (right) of SAO-CNN and the other NN-based in-loop filters proposed in [13] are compared for RA configuration. The proposed SAO-CNN framework can be considered as a good trade-off compared with the other low complexity methods.

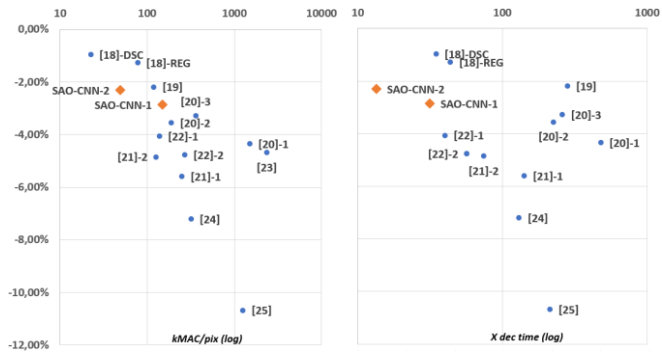

Fig. 5. BD-rate gain versus complexity (MAC/pel) and decoding time ratio with anchor of SAO-CNN and several proposals [13]:[19-25].

## V. CONCLUSION

The main purpose of this study was to study in-loop SAO post-filter based on CNN technique, targeting small complexity. The proposed approach, which combines multiple CNNs and integer inference, significantly improves the performance of the VTM while keeping complexity relatively small compared to other approaches. Previous works [16] have shown one could generally obtain better results with out-of-loop post-processing rather than with in-loop filtering, because the CNN-processed frames are employed as a reference (after in-loop filtering), what has not been reflected in the current CNN training, and it can lower the performance of other coding tools such as ALF. Another option to improve the efficiency of our method could be to reduce the number of CNNs by grouping video sequence classes or QPs at the training stage.

TABLE IV. SUMMARY OF ALL-INTRA CONFIGURATION BD-RATE GAIN OF THE PROPOSAED SAO-CNN COMPARED TO THE ANCHOR VTM-10 USING 32-FLOAT OR 16-BIT MODELS (V1, M=2).

| V1 | All Intra Main10 (32-float model) | | | Time | | All Intra Main10 (16-bit model) | | | Time | |
|---|---|---|---|---|---|---|---|---|---|---|
| Sequence | Y | U | V | EncT | DecT | Y | U | V | EncT | DecT |
| **Class A (2160p)** | -1.61% | -2.46% | -2.58% | 3.4× | 116.6× | -1.59% | -2.39% | -2.37% | 2.2× | 56.9× |
| **Class-B (1080p)** | -1.87% | -4.04% | -3.81% | 2.4× | 136.5× | -1.84% | -4.00% | -3.80% | 1.7× | 67.3× |
| **Class C (480p)** | -2.92% | -6.41% | -6.96% | 1.8× | 147.3× | -2.89% | -6.39% | -6.98% | 1.4× | 63.2× |
| **Class D (240p)** | -3.38% | -6.26% | -7.22% | 1.8× | 157.9× | -3.35% | -6.23% | -7.24% | 1.4× | 75.1× |
| **Overall** | -2.86% | -7.49% | -7.29% | 2.4× | 139.3× | -2.42% | -4.75% | -5.10% | 1.7× | 65.6× |

TABLE V. BD-RATE GAIN OF THE PROPOSAED SAO-CNN COMPARED TO THE ANCHOR VTM-10 USING 16-BIT MODEL (V1, M=2).

| Sequence | Random Access Main10 | | | Time | | All Intra Main10 | | | Time | |
|---|---|---|---|---|---|---|---|---|---|---|
| | Y | U | V | EncT | DecT | Y | U | V | EncT | DecT |
| A - Tango2 | -1.37% | -0.54% | -0.82% | 1.1× | 15.5× | -1.05% | -0.49% | -1.11% | 3.1× | 46.6× |
| A - FoodMarket4 | -1.01% | -2.18% | -2.37% | 1.2× | 8.2× | -1.78% | -1.06% | -1.21% | 2.7× | 56.6× |
| A - Campfire | -2.17% | -2.21% | -3.64% | 1.1× | 25.5× | -0.65% | -2.45% | -2.99% | 1.9× | 46.0× |
| A - CatRobot1 | -3.26% | -9.60% | -7.84% | 1.2× | 13.5× | -2.47% | -4.71% | -5.28% | 2.0× | 65.9× |
| A - DaylaightRoad2 | -4.18% | -8.73% | -6.15% | 1.2× | 12.8× | -1.47% | -3.64% | -2.22% | 2.0× | 58.6× |
| A - ParkRunning3 | -1.71% | -3.55% | -2.80% | 1.1× | 23.5× | -2.13% | -1.99% | -1.39% | 1.5× | 68.7× |
| **Class A (2160p)** | -2.28% | -4.47% | -3.94% | 1.2× | 16.5× | -1.59% | -2.39% | -2.37% | 2.2× | 56.9× |
| B - MarketPlace | -1.77% | -10.00% | -7.63% | 1.2× | 16.6× | -1.88% | -4.42% | -3.72% | 1.7× | 72.9× |
| B - RitualDance | -2.05% | -3.51% | -3.39% | 1.3× | 18.6× | -3.23% | -3.56% | -4.27% | 2.0× | 66.6× |
| B - Cactus | -2.75% | -6.06% | -5.51% | 1.3× | 20.3× | -1.93% | -1.75% | -0.69% | 1.6× | 69.5× |
| B - BasketballDrive | -1.92% | -4.28% | -4.85% | 1.2× | 22.9× | -1.26% | -5.24% | -6.76% | 1.8× | 69.1× |
| B - BQTerrace | -3.23% | -9.72% | -8.26% | 1.4× | 20.9× | -0.89% | -5.00% | -3.55% | 1.5× | 59.0× |
| **Class-B (1080p)** | -2.35% | -6.71% | -5.93% | 1.3× | 19.7× | -1.84% | -4.00% | -3.80% | 1.7× | 67.3× |
| C – BasketballDrill | -3.08% | -7.71% | -9.52% | 1.3× | 33.0× | -4.62% | -9.22% | -12.04% | 1.4× | 71.0× |
| C – BQMall | -3.52% | -9.93% | -9.99% | 1.3× | 29.2× | -3.25% | -6.48% | -6.94% | 1.4× | 66.4× |
| C – PartyScene | -2.86% | -9.92% | -6.79% | 1.4× | 46.7× | -2.19% | -5.81% | -3.82% | 1.2× | 48.8× |
| C – RaceHorses | -1.87% | -6.63% | -7.89% | 1.2× | 33.5× | -1.50% | -4.07% | -5.13% | 1.4× | 69.5× |
| **Class C (480p)** | -2.83% | -8.55% | -8.55% | 1.3× | 35.0× | -2.89% | -6.39% | -6.98% | 1.4× | 63.2× |
| D - BasketballPass | -3.33% | -10.52% | -10.91% | 1.3× | 58.9× | -3.97% | -7.15% | -9.63% | 1.5× | 90.7× |
| D – BQSquare | -6.39% | -9.75% | -13.34% | 1.6× | 73.1× | -3.18% | -4.20% | -6.75% | 1.3× | 68.0× |
| D – BlowingBubble | -3.01% | -9.04% | -7.00% | 1.3× | 41.4× | -2.70% | -5.28% | -4.03% | 1.3× | 64.7× |
| D - RaceHorses | -3.12% | -11.52% | -11.77% | 1.2× | 52.2× | -3.56% | -8.29% | -8.54% | 1.4× | 79.7× |
| **Class D (240p)** | -3.96% | -10.21% | -10.75% | 1.4× | 55.2× | -3.35% | -6.23% | -7.24% | 1.4× | 75.1× |
| **Overall** | -2.86% | -7.49% | -7.29% | 1.3× | 31.6× | -2.42% | -4.75% | -5.10% | 1.7× | 65.6× |

TABLE VI. SUMMARY OF BD-RATE GAIN OF THE PROPOSAED SAO-CNN COMPARED TO THE ANCHOR VTM-10 USING 16-BIT MODEL (M=2).

| V1 | Random Access Main10 | | | Time | | All Intra Main10 | | | Time | |
|---|---|---|---|---|---|---|---|---|---|---|
| Sequence | Y | U | V | EncT | DecT | Y | U | V | EncT | DecT |
| **Class A (2160p)** | -2.28% | -4.47% | -3.94% | 1.2× | 16.5× | -1.59% | -2.39% | -2.37% | 2.2× | 56.9× |
| **Class-B (1080p)** | -2.35% | -6.71% | -5.93% | 1.3× | 19.7× | -1.84% | -4.00% | -3.80% | 1.7× | 67.3× |
| **Class C (480p)** | -2.83% | -8.55% | -8.55% | 1.3× | 35.0× | -2.89% | -6.39% | -6.98% | 1.4× | 63.2× |
| **Class D (240p)** | -3.96% | -10.21% | -10.75% | 1.4× | 55.2× | -3.35% | -6.23% | -7.24% | 1.4× | 75.1× |
| **Overall** | -2.86% | -7.49% | -7.29% | 1.3× | 31.6× | -2.42% | -4.75% | -5.10% | 1.7× | 65.6× |

| V2 | Random Access Main10 | | | Time | | All Intra Main10 | | | Time | |
|---|---|---|---|---|---|---|---|---|---|---|
| Sequence | Y | U | V | EncT | DecT | Y | U | V | EncT | DecT |
| **Class A (2160p)** | -1.76% | -2.97% | -1.89% | 1.1× | 6.1× | -1.30% | -1.42% | -1.00% | 1.5× | 23.4× |
| **Class-B (1080p)** | -1.86% | -4.91% | -3.79% | 1.2× | 8.7× | -1.46% | -2.73% | -2.21% | 1.3× | 28.0× |
| **Class C (480p)** | -2.29% | -5.01% | -4.69% | 1.2× | 15.2× | -2.38% | -3.55% | -3.74% | 1.2× | 31.5× |
| **Class D (240p)** | -3.31% | -5.65% | -6.04% | 1.2× | 23.8× | -2.81% | -3.36% | -3.76% | 1.2× | 34.4× |
| **Overall** | -2.31% | -4.64% | -4.10% | 1.2× | 13.5× | -1.99% | -2.77% | -2.68% | 1.3× | 29.3× |

REFERENCES

[1] V.Baroncini, J-R.Ohm and G.Sullivan, "Report of results from the Call for Proposals on Video Compression with Capability beyond HEVC," document JVET-J1003, 10th Meeting: San Diego, US, 10–20 Apr. 2018.

[2] B. Bross, J. Chen, S. Liu, Y.-K. Wang , "Versatile Video Coding Editorial Refinements on Draft 10," document JVET-T2001, 20th JVET Meeting, by teleconference, 7 – 16 Oct. 2020.

[3] A/Norkin, G.Bjontegaard, A.Fuldseth, M.Narroschke, M.Ikeda, K.Andersson, M.Zhou and G.Van der Auwera, "HEVC Deblocking Filter," IEEE Transactions on Crircuits and Systems for Video Technology, Vol. 22, No. 12, Dec. 2012.

[4] C.Fu, E.Alshina, A.Alshin, Y.Huang, C.Chen and C.Tsai, "Sample Adaptive Offset in the HEVC Standard," IEEE Transactions on Circuits and Systemes for Video Technology, Vol. 22, No. 12, Dec. 2012.

[5] C.Tsai, C.Chen, T.Yamakage, I.Chong, Y.Huang, C.Fu, T.Itoh, T.Watanabe, T.Chujoh, M.Karczewicz and S.Lei, "Adaptive Loop Filtering for Video Coding," IEEE Journal of Selected Topics in Signal Processing, Vol. 7, NO. 6, Dec. 2013.

[6] W.-J. Chien, J. Boyce, Y.-W. Chen, R. Chernyak, K. Choi, R. Hashimoto, Y.-W. Huang, H. Jang, R.-L. Liao, S. Liu , "JVET AHG report: Tool reporting procedure and testing (AHG13)," document JVET-T0013, 20th JVET Meeting by teleconference, 7 – 16 Oct. 2020.

[7] F.Bossen, X.Li, K.Suehring, "AHG report: Test model software development (AHG3)," document JVET-T0003, 20th JVET Meeting: by teleconference, 7 – 16 Oct. 2020.

[8] F.Bossen, K.Suehring, X.Li, V.Seregin, "VTM common test conditions and software reference configurations for SDR video," document JVET-T2010, 20th Meeting, by teleconference, 7 – 16 Oct. 2020.

[9] G.Bjøntegaard, "Improvements of the BD-PSNR Model," document VCEG-AI11, ITU-T SG 16/Q6, 35th VCEG Meeting, Jul. 2008.

[10] W.-J. Chien, J. Boyce, Y.-W. Chen, R. Chernyak, K. Choi, R. Hashimoto, Y.-W. Huang, H. Jang, R.-L. Liao, S. Liu , "JVET AHG report: Tool reporting procedure and testing (AHG13)," document JVET-T0013, 20th JVET Meeting, by teleconference, 7 – 16 Oct. 2020.

[11] E.François, M.Kerdranvat, R.Julian, C.Chevance, P.DeLagrange, F.Urban, T.Poirier, Y.Chen, "VVC per-tool performance evaluation compared to HEVC," IBC Sept. 2020.

[12] C.Fu, C.Chen, Y.Huang, S.Lei, "CE8 Subset3: Picture Quadtree Adaptive Offset," document JCTVC-D122, 4th JCTVC Meeting Jan. 2011.

[13] S.Liu, A.Segall, Y.Ye, E.Alshina, J.Chen, F.Galpin, J.Pfaff, S.Wang, M.Wien, P.Wu, J.Xu, "JVET AHG report: Neural Network-based video coding," document JVET-U0011, 21th JVET Meeting, Jan. 2021.

[14] Y.Sun, N.B.Agostini, S.Dong and D.Kaeli, "Summarizing CPU and GPU Design Trends with Product Data," https://arxiv.org/pdf/1911.11313.pdf .

[15] L.Kong, D.Ding, F.Liu, D.Mukherjee, U.Joshi, Y.Chen, "Guided CNN Restoration with Explicitly Signaled Linear Combination," IEEE International Conference on Image Processing (ICIP), Oct. 2020.

[16] D.Ma, F.Zhang, D.R.Bull, "MFRNet: A New CNN Architecture for Post-Processing and In-loop Filtering," https://arxiv.org/abs/2007.07099 , Jul. 2020.

[17] S.Lee, Y.Yang, S.Cho, B.Oh, "Offset-based in-Loop Filtering with a Deep Network in HEVC," IEEE Access, Dec. 2020.

[18] D. Ma, F. Zhang, and D. R. Bull, "BVI-DVC: a training database for deep video compression," arXiv preprint arXiv:2003.13552, 2020

[19] C.Auyeung, W.Wei, W.Jiang, X.Li, S.Liu, "EE1.1: A comparison of depthwise separable convolution and regular convolution with the JVET-T0057 neural network based in-loop filter," doc. JVET-U0060, 21st Meeting, by teleconference, 6–15 Jan. 2021.

[20] T.Ouyang, H.Zhu, Z.Chen, X.Xu, S.Liu, "EE: SSIM based CNN model for in-loop filtering," doc. JVET-U0074, 21st Meeting, by teleconference, 6–15 Jan. 2021.

[21] W.Chen, X.Xiu, Y.Chen, H.Jhu, C.Kuo, X.Wang, "EE-2.1.5: In-loop filtering based on neural network," doc. JVET-U0101, 21st Meeting, by teleconference, 6–15 Jan. 2021.

[22] H.Wang, M.Karczewicz, J.Chen, A.Kotra, "EE: Tests on Neural Network-based In-Loop Filter," doc. JVET-U0094, 21st Meeting, by teleconference, 6–15 Jan. 2021.

[23] J.Chen, H.Wang, A.Kotra, M.Karczewicz, "AHG11: In-loop filtering with convolutional neural network and large activation," dic. JVET-U0104, 21th Meeting: by teleconference, 6 – 15 Jan 2021.

[24] Z.Wang, R.Liao, C.Ma, Y.Ye, "EE-1.6: Neural network based in-loop filtering," dic. JVET-U0054, 21st Meeting, by teleconference, 6–15 Jan. 2021.

[25] H.Wang, J.Chen, A.Kotra, M.Karczewicz, "AHG11: Neural Network-based In-Loop Filter Performance with No Deblocking Filtering stage," doc. JVET-U0115, 21st Meeting, by teleconference, 6–15 Jan. 2021.

[26] Y.Li, L.Zhang, K.Zhang, "AHG11: Convolutional Neural Network-based In-Loop Filter with Adaptive Model Selection," doc. U0068, 21st Meeting, by teleconference, 6–15 Jan. 2021.